\documentclass[preprint,prd,nofootinbib,tightenlines,amsmath]{revtex4}
\usepackage{graphicx}
\usepackage{bm}
\usepackage{color}

\begin{document}
\baselineskip=15pt \parskip=5pt

\vspace*{3em}

\title{Constraints on Scalar Dark Matter from Direct Experimental Searches}

\author{Xiao-Gang He$^{1}$}
\email{hexg@phys.ntu.edu.tw}
\author{Tong Li$^{2,3}$}
\email{allongde@mail.nankai.edu.cn}
\author{Xue-Qian Li$^{2}$}
\email{lixq@nankai.edu.cn}
\author{Jusak Tandean$^{1}$}
\email{jtandean@yahoo.com}
\author{Ho-Chin Tsai$^{1}$}
\email{hctsai@phys.ntu.edu.tw}

\affiliation{$^1$Department of Physics, Center for Theoretical Sciences, and LeCosPA Center, \\
National Taiwan University, Taipei, Taiwan \\
$^2$Department of Physics, Nankai University, Tianjin 300071, China \\
$^3$Center for High Energy Physics, Peking University, Beijing 100871, China}

\date{\today $\vphantom{\bigg|_{\bigg|}^|}$}

\begin{abstract}
The standard model (SM) plus a real gauge-singlet scalar field dubbed darkon (SM+D) is the
simplest model possessing a weakly interacting massive particle (WIMP) dark-matter candidate.
The upper limits for the WIMP-nucleon elastic cross-section as a function of WIMP mass from the
recent XENON10 and CDMS\,II experiments rule out darkon mass ranges from 10 to \,$(50,70,75)$\,GeV
for Higgs-boson masses of \,$(120,200,350)$\,GeV,\, respectively.  This may exclude the possibility
of the darkon providing an explanation for the gamma-ray excess observed in the EGRET data.
We show that by extending the SM+D to a~two-Higgs-doublet model plus a darkon the experimental
constraints on the WIMP-nucleon interactions can be circumvented due to suppression occurring at
some values of the product  \,$\tan\alpha\,\tan\beta$,\, with $\alpha$ being the neutral-Higgs
mixing angle and $\tan\beta$ the ratio of vacuum expectation values of the Higgs doublets.
We also comment on the implication of the darkon model for Higgs searches at the~LHC.
\end{abstract}

\maketitle

\section{Introduction}

The standard big-bang cosmology describes various astronomical and cosmological observations very
well.  To explain observational data, it requires two exotic unknown species beyond the standard
model~(SM) of particle physics, namely the dark matter and the dark energy, which in the energy
budget of the Universe make up roughly 20\% and 75\% of the total energy density, respectively.
Although the evidence for dark matter has been established for many decades, the identity of
its basic constituents has so far remained elusive.

\begin{figure}[b]
\includegraphics[height=2.7in,width=3in]{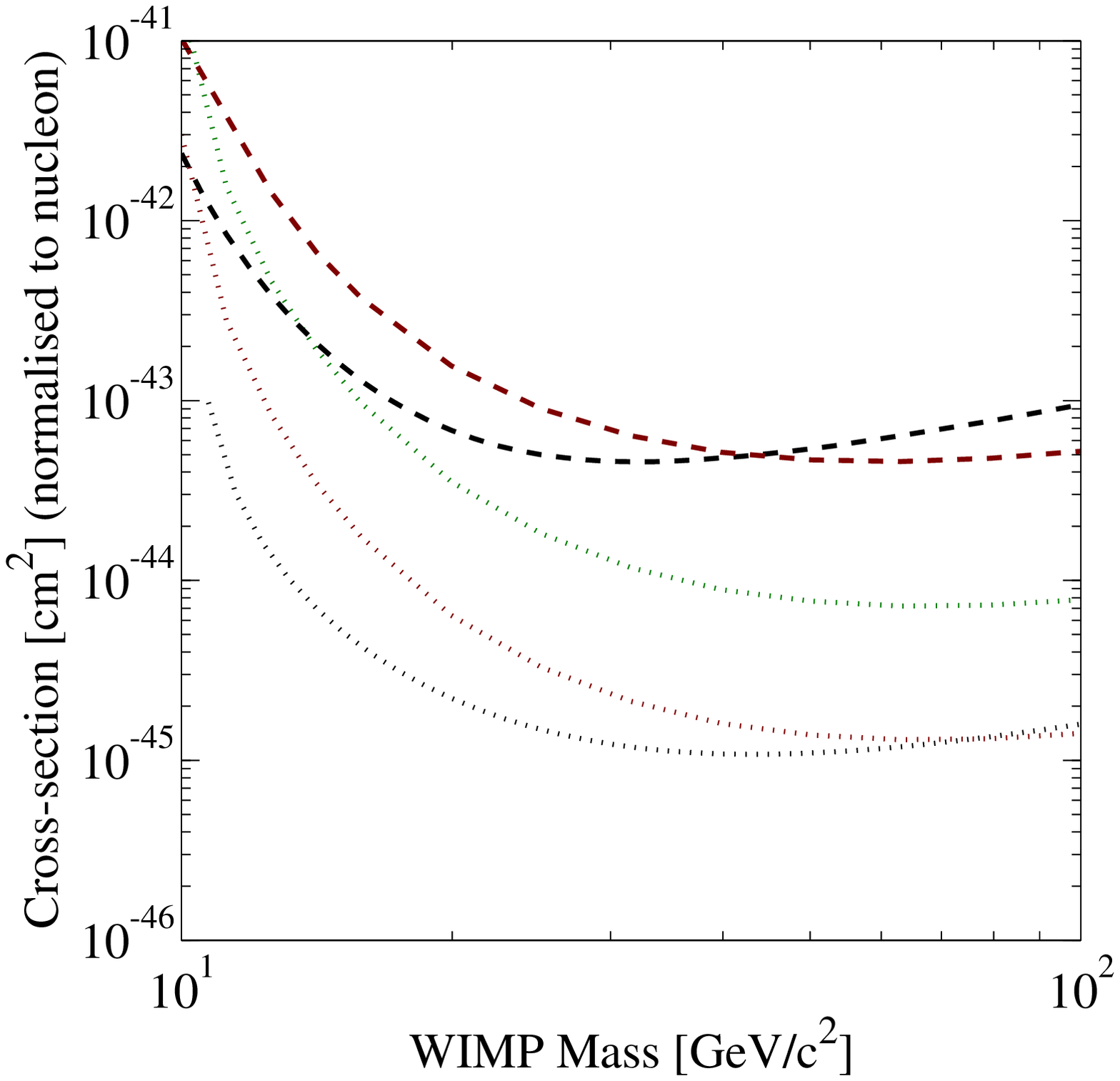}
\includegraphics[width=3in]{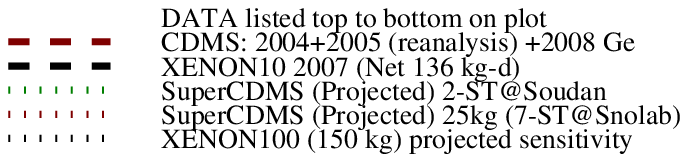}
\caption{Current and projected upper limits for the spin-independent WIMP-nucleon elastic
scattering cross-section as functions of WIMP mass.
Plot courtesy of Ref.~\cite{dmplot}.
\label{dm_search}}
\end{figure}

One of the popular candidates for dark matter (DM) is the weakly interacting massive particle
(WIMP).  Needless to say, the detection of a WIMP candidate is crucial not only for
understanding the nature of DM, but also for distinguishing models of new physics beyond the SM.
A~variety of experiments have been and are being carried out to detect DM directly by looking
for the recoil energy of nuclei caused by the elastic scattering of a WIMP off a nucleon.
Although there is some evidence for WIMPs from indirect DM searches~\cite{deBoer:2008iu},
there is as yet no confirmed signal for them from direct searches.
Stringent bounds on the WIMP-nucleon elastic cross-section have been obtained from
the null results of direct DM searches.  The strictest limits to date on the cross section as
a~function of WIMP mass have been set by the recent XENON10~\cite{Angle:2007uj} and
CDMS\,II~\cite{Ahmed:2008eu} experiments.  They are shown in Fig.~\ref{dm_search}, along with
the expected sensitivities of a few future experiments.
These upper bounds provide important restrictions on WIMP models.

Among the many possible WIMPs, the lightest supersymmetric particle is the most studied
candidate.  Although this possibility has many attractive features, no direct experimental
evidence for it has yet been discovered.  It is therefore important to study and search for
other possible WIMPs which can explain the DM relic density.
The simplest model which has a WIMP candidate is the SM+D, which extends the SM by the addition
of a real gauge-singlet scalar field $D$.  This singlet field, which we will call darkon, can
play the role of dark matter.
The darkon as DM was first considered by Silveira and Zee~\cite{Silveira:1985rk} and further
explored later by other groups~\cite{Burgess:2000yq,darkon,He:2007tt,Barger:2007im,Yaguna:2008hd}.

In this paper we study the constraints on the SM+D and its two-Higgs-doublet extension (THDM+D)
from the recent XENON10 and CDMS\,II experiments which set upper limits for the WIMP-nucleon
spin-independent elastic scattering cross-section~\cite{Angle:2007uj,Ahmed:2008eu}.
In the SM+D, we find that a darkon having a mass in the ranges of 10 GeV to $(50, 70, 75)$\,GeV
for Higgs masses of $(120, 200, 350)$\,GeV, respectively, is ruled out by the experimental bounds.
This confirms some of the results obtained in Refs.~\cite{Barger:2007im,Yaguna:2008hd}.
Now, a WIMP candidate with a mass in the range from 50 to 70\,GeV can well explain
the gamma-ray excess observed in the EGRET data~\cite{deBoer:2005tm}.
The darkon mass range ruled out by XENON10 and CDMS\,II would imply that the darkon model is
not likely to offer an explanation for the EGRET excess.
It is then interesting to see if there are possible avenues to evade the experimental
restrictions on the WIMP-nucleon interactions and keep this darkon mass range viable.
To this end, we consider two-Higgs-doublet models with a darkon and find that in
the type-II THDM plus a darkon (THDM\,II+D) it is indeed possible for a~darkon with a mass
within the range of interest to satisfy the experimental requirements.
This can happen because of the suppression of the darkon-nucleon interaction at certain values
of the parameters of the model.
We also find that a darkon having mass within this range can lead to considerable enhancement
of Higgs decay widths through substantial invisible decays into darkon pairs.
This could cause the Higgs branching fractions to SM particles to diminish and consequently
affect Higgs signatures at colliders significantly.

Before starting our analysis of the models, we would like to summarize the relic-density
constraints that any WIMP candidate has to satisfy.
For a given interaction of the WIMP with SM particles, its annihilation rate into the latter and
its relic density $\Omega_D^{}$ can be calculated and are related to each other by the thermal
dynamics of the Universe within the standard big-bang cosmology~\cite{Kolb:1990vq}.
To a good approximation,
\begin{eqnarray} \label{oh}
\Omega_D^{} h^2 \,\,\simeq\,\,
\frac{1.07\times 10^9\, x_f^{}}{
\sqrt{g_*^{}}\, m_{\rm Pl}\,\langle\sigma_{\rm ann}^{}v_{\rm rel}^{}\rangle\rm\,GeV} \,\,,
\hspace{2em}
x_f^{} \,\,\simeq\,\,
\ln\frac{0.038\,m_{\rm Pl}\,m_D^{}\,\langle\sigma_{\rm ann}^{}v_{\rm rel}^{}\rangle}{
\sqrt{g_*^{}\, x_f^{}}} \,\,,
\end{eqnarray}
where  $h$ is the Hubble constant in units of 100\,km/(s$\cdot$Mpc),
\,$m_{\rm Pl}^{}=1.22\times10^{19}$\,GeV\, is the Planck mass, $m_D^{}$ is the WIMP mass,
\,$x_f^{}=m_D^{}/T_f^{}$\, with $T_f^{}$ being the freezing temperature, $g_*^{}$ is the number
of relativistic degrees of freedom with masses less than $T_f^{}$, and
\,$\langle\sigma_{\rm ann}^{}v_{\rm rel}^{}\rangle$\, is the thermal average of the product of
the annihilation cross-section of a pair of WIMPs into SM particles and the relative speed of
the WIMP pair in their center-of-mass frame.

\begin{figure}[th]
\includegraphics[height=2in,width=3in]{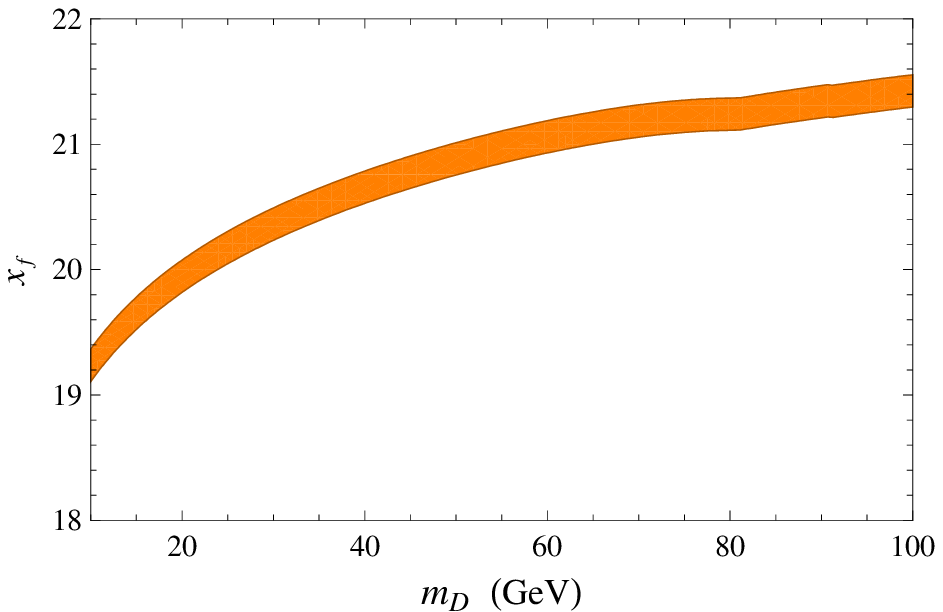} \, \, \,
\includegraphics[height=2in,width=3in]{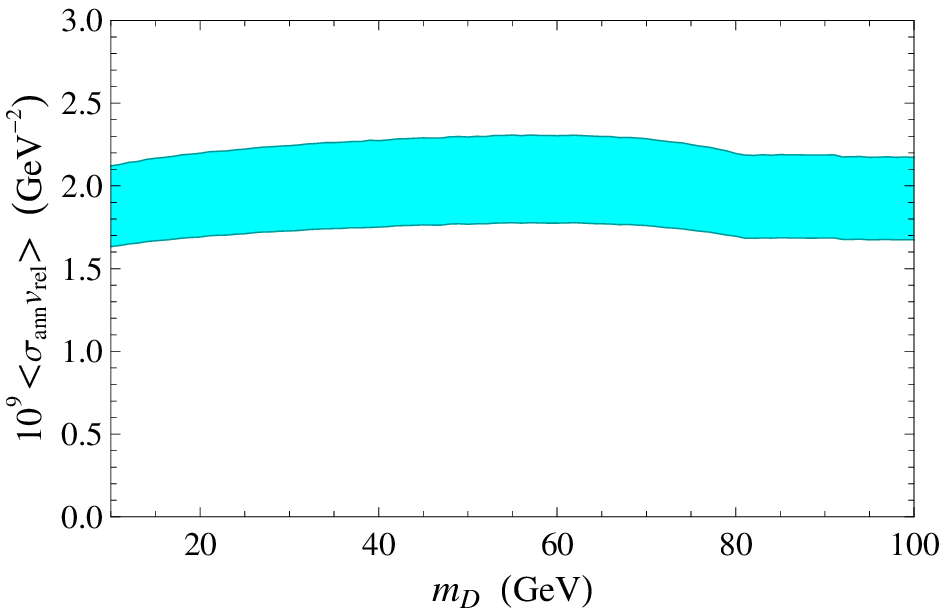}
\caption{Ranges of $x_f^{}$ and  \,$\langle\sigma_{\rm ann}^{}v_{\rm rel}^{}\rangle$\, as functions
of WIMP mass $m_D$ corresponding to the 90\%\,C.L. range \,$0.092\le\Omega_D^{}h^2\le0.118$.
\protect\label{xfsigma}}
\end{figure}

The current Particle Data Group value for the DM density is
\,$\Omega_D^{} h^2=0.105\pm0.008$\,~\cite{Amsler:2008zz}.
Using this number and Eq.~(\ref{oh}), one can restrict the ranges of $x_f^{}$ and
\,$\langle\sigma_{\rm ann}^{}v_{\rm rel}^{}\rangle$\,  as functions of WIMP mass $m_D^{}$
without knowing the explicit form of the SM-WIMP interaction.
In Fig.~\ref{xfsigma} we show the values allowed by the 90\%\,C.L. range
\,$0.092\le\Omega_D^{}h^2\le0.118$.

\section{Standard model with darkon\label{sm+d}}

\subsection{Brief description of the model}

Since the darkon field $D$ in the SM+D must interact weakly with the SM matter fields to play
the role of dark matter, the simplest way to introduce the darkon is to make it a real gauge
singlet with a discrete $Z_2$ symmetry so that it does not have SM gauge interactions and
can only be created or annihilated in pairs.  Requiring that the darkon interactions be
renormalizable implies that $D$ can only couple to the Higgs doublet field $H$.
Beside the kinetic energy term  \,$\frac{1}{2}\partial^\mu D\,\partial_\mu^{}D$,\, the general
form of the other terms in the darkon Lagrangian is~\cite{Silveira:1985rk,Burgess:2000yq}
\begin{eqnarray}  \label{DH}
{\cal L}_D^{} \,\,=\,\, -\frac{\lambda_D^{}}{4}\,D^4
- \frac{m_0^2}{2}\,D^2 - \lambda\, D^2\,H^\dagger H \,\,,
\end{eqnarray}
where  $\lambda_D^{}$,  $m_0^{}$, and $\lambda$  are free parameters.  Clearly, ${\cal L}_D^{}$
is invariant under the $Z_2$ transformation \,$D\to-D$,\, the SM fields being unchanged.
The parameters in the potential should be chosen such that $D$ does not develop a vacuum
expectation value (vev) and the $Z_2$ symmetry is not broken, which will ensure that the darkon
does not mix with the Higgs field, avoiding possible fast decays into other SM particles.

The Lagrangian in Eq.~(\ref{DH}) can be rewritten to describe the interaction of the physical
Higgs boson $h$ with the darkon as
\begin{eqnarray}
{\cal L}_D^{} \,\,=\,\, -\frac{\lambda_D^{}}{4}\,D^4-\frac{\bigl(m_0^2+\lambda v^2\bigr)}{2}\,D^2
- \frac{\lambda}{2}\, D^2\, h^2 - \lambda v\, D^2\, h \,\,,
\end{eqnarray}
where  \,$v=246$\,GeV\,  is the vev of $H$,  the second term contains the darkon mass
\,$m_D^{}=\bigl(m^2_0+\lambda v^2\bigr)^{1/2}$,\, and the last term,  \,$-\lambda v D^2 h$,\,
plays an important role in determining the relic density of the DM.
At leading order, the relic density of the darkon results from the annihilation of a darkon
pair into SM particles through Higgs exchange~\cite{Silveira:1985rk,Burgess:2000yq}, namely
\,$DD\to h^*\to X$,\, where $X$ indicates SM particles.

Since the darkon is cold DM, its speed is nonrelativistic, and so a darkon pair has an invariant
mass  \,$\sqrt s\simeq2m_D^{}$.\,  With the SM+D Lagrangian determined, the $h$-mediated
annihilation cross-section of a darkon pair into SM particles is then given by~\cite{Burgess:2000yq}
\begin{eqnarray} \label{csan}
\sigma_{\rm ann}^{}\, v_{\rm rel}^{} \,\,=\,\,
\frac{8\lambda^2 v^2}{\bigl(4m_D^2-m_h^2\bigr)^2+\Gamma^2_h\,m^2_h}\,
\frac{\sum_i\Gamma\bigl(\tilde h\to X_i^{}\bigr)}{2m_D^{}} \,\,,
\end{eqnarray}
where  \,$v_{\rm rel}^{}=2\bigl|\bm{p}_D^{\rm cm}\bigr|/m_D^{}$\, is the relative speed of
the $DD$ pair in their center-of-mass (cm) frame, $\tilde h$  is a~virtual Higgs boson having
the same couplings to other states as the physical $h$ of mass $m_h^{}$, but with an invariant mass
\,$\sqrt s=2m_D^{}$, and \,$\tilde h\to X_i$\, is any possible decay mode of $\tilde h$.
For a given model, \,$\Sigma_i\Gamma\bigl(\tilde h\to X_i\bigr)$\, is obtained by calculating
the $h$ width and then setting $m_h^{}$ equal to  $2m_D^{}$.

\begin{figure}[hb]
\includegraphics[width=3.5in]{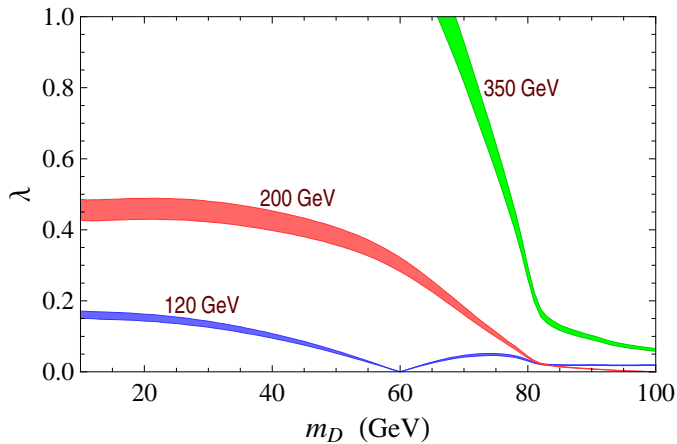}
\caption{\label{lambda_sm}%
Darkon-Higgs coupling $\lambda$ in the SM+D as a function of darkon mass $m_D^{}$
for Higgs mass values  \,$m_h^{}=120,200,350$\,GeV.}
\end{figure}

The darkon-Higgs coupling $\lambda$ for a given value of $m_D^{}$ can now be inferred from
the range of \,$\langle\sigma_{\rm ann}^{}v_{\rm rel}^{}\rangle$\, values allowed by the
$\Omega_D^{}h^2$ constraint, as in Fig.~\ref{xfsigma}, once $m_h^{}$ is specified.
We show in Fig.~\ref{lambda_sm} the allowed ranges of  $\lambda$  corresponding to
\,$10{\rm\,GeV}\le m_D^{}\le100{\rm\,GeV}$\,  for representative values of the Higgs-boson mass.
We note that it is possible for $\lambda$ to become larger than 1 when $m_D^{}$
decreases, which would upset the applicability of perturbative calculation.
Consequently, we display only the \,$\lambda<1$\, regions in this figure.
\\

\subsection{Effective Higgs-nucleon coupling}

The detection of dark matter on the Earth is through the recoil of nuclei when a darkon
hits a~nucleon target.  This interaction occurs via the exchange of a Higgs boson between
the darkon and the nucleon $N$ in the $t$-channel process \,$DN\to DN$.\,
Its cross-section then depends on not only $\lambda$, but also the Higgs-nucleon coupling.
Since the energy transferred in this elastic scattering is very small, of order 100\,keV,
one can employ a chiral-Lagrangian approach to obtain the effective Higgs-nucleon coupling.
The Higgs-nucleon coupling has been studied previously~\cite{Shifman:1978zn,Cheng:1988cz},
and here we reorganize the results for convenience in evaluating possible cancelation among
different contributions.

The Higgs-nucleon coupling depends on the underlying Yukawa interactions of the Higgs boson
with the quark degrees of freedom.  The couplings of a Higgs boson $\cal H$ to quarks can be
generically written as
\begin{eqnarray}
{\cal L}_{qq\cal H}^{} \,\,=\,\, -\sum_q\frac{k_q^{}}{v}\, m_q^{}\, \bar q q\, \cal H \,\,,
\end{eqnarray}
where the sum runs over the six quark flavors, \,$q=u,d,s,c,b,t$.\,  In the SM \,$k_q^{}=1$\,
for all $q$'s, whereas beyond the SM the $k_q^{}$'s may have other values.
Specifically, \,$k_u^{}=k_c^{}=k_t^{}$\,  and  \,$k_d^{}=k_s^{}=k_b^{}$,\, the two sets
being generally different,  in the THDM\,II+D
which we will consider later.

Now, the effective coupling of $\cal H$ to a nucleon \,$N=p$ or $n$\, has the form
\begin{eqnarray}
{\cal L}_{NN\cal H}^{} \,\,=\,\, -g_{NN\cal H}^{}\, \bar N N\, {\cal H} \,\,,
\end{eqnarray}
where  $g_{NN\cal H}^{}$ is the Higgs-nucleon coupling constant.
In the SM+D and THDM\,II+D, one then needs to evaluate the matrix element
\begin{eqnarray} \label{gnnh0}
g_{NN\cal H}^{}\, \bar N N \,\,=\,\,
\langle N| \frac{k_u^{}}{v} (m_u^{}\, \bar u u+m_c^{}\, \bar c c+m_t^{}\, \bar t t)+
\frac{k_d^{}}{v}( m_d^{}\, \bar d d+m_s^{}\, \bar s s+m_b^{}\, \bar b b )|N\rangle \,\,.
\end{eqnarray}
We outline the derivation in the Appendix, and the result is
\begin{eqnarray} \label{gNNH}
g_{NN\cal H}^{} \,\,=\,\,
\bigl(k_u^{}-k_d^{}\bigr) \frac{\sigma_{\pi N}^{}}{2 v} \,+\,
k_d^{}\, \frac{m_N^{}}{v} \,+\, \frac{4k_u^{}-25k_d^{}}{27}\,\, \frac{m_B^{}}{v}  \,\,,
\end{eqnarray}
where $\sigma_{\pi N}^{}$ is the so-called pion-nucleon sigma term, $m_N^{}$ the nucleon mass,
and $m_B^{}$ the baryon mass in the chiral limit.  As explained in the Appendix, $m_B^{}$ is
related to $\sigma_{\pi N}^{}$ by Eq.~(\ref{m0}) and numerically we adopt
\,$\sigma_{\pi N}^{}=45$\,MeV.\,  It follows that
\begin{eqnarray} \label{g_nnh}
g_{NN\cal H}^{} \,\,\simeq\,\, \bigl(1.217\,k_d^{}+0.493\, k_u^{}\bigr)\times10^{-3} \,\,.
\end{eqnarray}
Hence  $g_{NN\cal H}^{}$ can be vanishingly small if  \,$k_u^{}\simeq-2.47\,k_d^{}$.

\subsection{Darkon-nucleon elastic cross-section in SM+D}

The amplitude for the elastic scattering \,$DN\to DN$\, mediated by $\cal H$ is given by
\begin{eqnarray}
{\cal M}_{\rm el}^{} \,\,\simeq\,\, \frac{2\lambda\,g_{NN\cal H}^{}\,v}{m_{\cal H}^2}\,\bar NN
\end{eqnarray}
for the squared transfer momentum \,$t\ll m_{\cal H}^2$.\,
This leads to the cross section
\begin{eqnarray}
\sigma_{\rm el}^{} \,\,\simeq\,\,
\frac{\lambda^2\,g_{NN\cal H}^2\,v^2\,m_N^2}{\pi\,\bigl(m_D^{}+m_N^{}\bigr)^2\, m_{\cal H}^4} \,\,,
\end{eqnarray}
the approximation  \,$\bigl(p_D^{}+p_N^{}\bigr)^2\simeq\bigl(m_D^{}+m_N^{}\bigr)^2$\, having
been used.
In the SM+D, the Higgs-nucleon coupling $g_{NNh}^{\rm SM}$ is obtained from Eq.~(\ref{g_nnh}) by
setting  \,$k_u^{}=k_d^{}=1$.\,  Thus
\begin{eqnarray}
g_{NNh}^{\rm SM} \,\,\simeq\,\, 1.71\times10^{-3}  \,\,,
\end{eqnarray}
which is comparable to the values found in the literature~\cite{Burgess:2000yq,Cheng:1988cz}.

\begin{figure}[ht]
\includegraphics[width=3.5in]{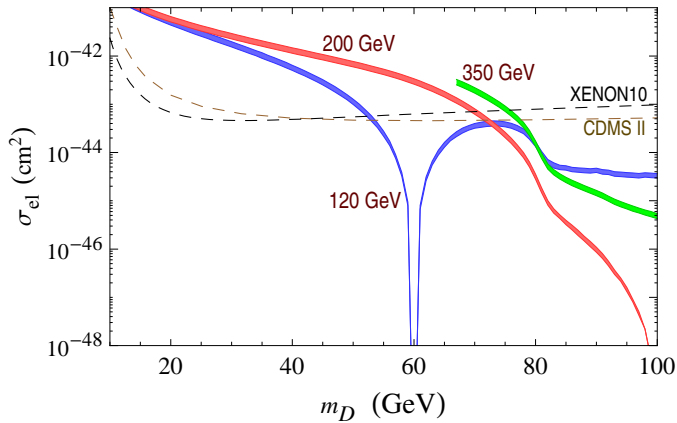}
\caption{Darkon-nucleon elastic cross-section $\sigma_{\rm el}^{}$ in the SM+D as a function of
darkon mass $m_D^{}$ for Higgs mass values \,$m_h^{}=120,200,350$\,GeV,\, compared to
90\%\,C.L. upper limits from XENON10 (black dashed-curve) and CDMS\,II (gray [brown] dashed-curve).
\label{elastic_smd}}
\end{figure}

Since $\lambda$ as a function of $m_D^{}$ is constrained by the relic density, so is
$\sigma_{\rm el}^{}$.  In Fig.~\ref{elastic_smd} we plot $\sigma_{\rm el}^{}$  versus $m_D^{}$
in the SM+D  for  \,$\lambda<1$\,  and  representative values of the Higgs-boson mass.
We remark that the dip of the \,$m_h^{}=120$\,GeV curve at \,$m_D^{}=60$\,GeV\, in this figure
corresponds to the dip in Fig.~\ref{lambda_sm} and the minimum of the denominator in
Eq.~(\ref{csan}) at \,$m_h^{}=2m_D^{}$.\,
When comparing predicted elastic cross-sections with experimental limits, we assume in this
paper that the local density of dark matter is fully accounted for by the darkon, and
so no scaling is needed for the cross sections.
In Fig.~\ref{elastic_smd}, we also plot 90\%\,C.L. upper limits on the WIMP-nucleon spin-independent
elastic cross-section set by the XENON10 and CDMS\,II experiments~\cite{Angle:2007uj,Ahmed:2008eu}.
One can easily see that the darkon mass ranges \,$10{\rm\,GeV}\le m_D^{}\le(50,70,75)$\,GeV\,
are ruled out for Higgs masses of \,$(120, 200, 350)$\,GeV,\, respectively.

The analysis above suggests that the experimental constraints can be evaded if the Higgs-nucleon
coupling gets sufficiently small due to (at least) partial cancelation among various contributions
to it from the underlying Yukawa couplings.
As we will see in the next section, this possibility can be realized in the two-Higgs-doublet
extension of the SM+D.
\\

\section{Two-Higgs-doublet model with darkon}

\subsection{Brief description of the model}

As the name implies, the THDM has two Higgs doublets $H_{1,2}$.  Depending on how they couple
to the fermions in the SM, there are different types of the THDM~\cite{thdm}.  In the THDM\,I,
only one of the Higgs doublets gives masses to all the fermions.
In the THDM\,II, the up-type fermions get mass from only one of the Higgs doublets, say $H_2$,
and the down-type fermions from the other doublet.
In the THDM\,III, both $H_1$ and $H_2$ give masses to all the fermions.

In the THDM\,I, since only one Higgs doublet generates the fermion masses, the Higgs couplings
to fermions are those in the SM, up to an overall scaling factor.
Therefore, the Higgs couplings in the THDM\,I+D are similar to those in the SM+D studied in
the previous section and thus cannot help overcome the difficulties encountered.
In the THDM\,III, there are flavor-changing Higgs-quark couplings which introduce too many
parameters for the model to have predictive power.
For these reasons, we will consider only the THDM\,II with the darkon field added (THDM\,II+D).

The Yukawa interactions of the Higgs fields in the THDM\,II are described by~\cite{thdm}
\begin{eqnarray} \label{yukawa_2hdm}
{\cal L}_{\rm Y}^{} \,\,=\,\, - \bar Q_L^{} \lambda^u_2 \tilde H_2^{}{\cal U}_R^{}
- \bar Q_L^{}\lambda^d_1 H_1^{} {\cal D}_R^{}
- \bar L_L^{} \lambda^l_1  H_1^{} E_R^{} \,\,+\,\, {\rm H.c.} \,\,,
\end{eqnarray}
where  $Q$, $\cal U$, $\cal D$, $L$, and $E$ represent the usual quark and lepton
fields and $\lambda^{u,d,l}$ contain the Yukawa couplings.
To effect the separate couplings of $H_1^{}$ and $H_2^{}$ to the down and up sectors, respectively,
it is necessary to introduce a discrete $Z'_2$ symmetry, under which \,$H_2^{}\to-H_2^{}$\, and
\,${\cal U}_R^{}\to-{\cal U}_R^{}$,\, the other fields being unaltered.
In terms of their components, the Higgs doublets are
\begin{eqnarray}
H_k^{} \,\,=\,\, \frac{1}{\sqrt2} \left(\begin{array}{c} \sqrt{2} h^+_k\\
v_k^{}+ h_k^{} + i I_k^{} \end{array}\right ) \,\,,
\end{eqnarray}
where \,$k=1,2$\, and $v_k^{}$ is the vev of $H_k^{}$.
Here $h^+_k$ and $I_k^{}$ are related to the physical Higgs bosons $H^+$ and $A$ and
the would-be Goldstone bosons $w$ and $z$ by
\begin{eqnarray}
\left(\begin{array}{c}h^+_1 \\ h^+_2\end{array}\right) \,\,=\,\,
\left(\begin{array}{rrr} \cos\beta && -\sin\beta \\ \sin\beta && \cos\beta \end{array}\right)
\left(\begin{array}{c}w^+\\H^+\end{array}\right) \,\,, \hspace{2em}
\left(\begin{array}{c}I_1^{} \\ I_2^{} \end{array}\right) \,\,=\,\,
\left(\begin{array}{rrr}\cos\beta &&-\sin\beta \\ \sin\beta && \cos\beta\end{array}\right)
\left(\begin{array}{c}z\\A\end{array}\right) \,\,\,,
\end{eqnarray}
with  \,$\tan\beta=v_2^{}/v_1^{}$,\, whereas $h_k^{}$ can be expressed in terms of mass
eigenstates $H$ and $h$ as
\begin{eqnarray}
\left(\begin{array}{c}h_1^{} \\ h_2^{} \end{array}\right) \,\,=\,\,
\left(\begin{array}{rrr} \cos\alpha && -\sin\alpha \\ \sin\alpha && \cos\alpha\end{array}\right)
\left(\begin{array}{c}H \\ h \end{array}\right) \,\,.
\end{eqnarray}
Hence the angle $\alpha$ indicates the mixing of the two $CP$-even Higgs bosons.

In analogy to Eq.~(\ref{DH}) in the SM+D case,  in the THDM\,II+D  we have the renormalizable
darkon Lagrangian
\begin{eqnarray}
{\cal L}_D^{} \,\,=\,\, -\frac{\lambda_D^{}}{4}\, D^4 - \frac{m_0^2}{2}\, D^2
- \bigl(\lambda_1^{}\, H_1^{\dag}H_1^{} + \lambda_2^{}\, H^\dagger_2 H_2^{} \bigr) D^2 \,\,.
\end{eqnarray}
As in the SM+D, here we have again imposed the $Z_2$ symmetry under which $D\to -D$
with all the other fields unchanged, and for the same reasons we need to keep it unbroken.
This Lagrangian also respects the $Z'_2$ symmetry mentioned in the preceding paragraph.

After electroweak symmetry breaking, ${\cal L}_D$ contains the $D$ mass and the $DD(h,H)$
terms given by
\begin{eqnarray} \label{lambda}
\begin{array}{c}   \displaystyle
m^2_D \,\,=\,\, m^2_0 + \bigl(\lambda_1^{}\,\cos^2\beta+\lambda_2^{}\,\sin^2\beta
\bigr)v^2 \,\,,
\vspace{1.5ex} \\   \displaystyle
{\cal L}_{DDh}^{} \,\,=\,\, -\bigl(-\lambda_1^{}\,\sin\alpha\,\cos\beta
+ \lambda_2^{}\,\cos\alpha\, \sin\beta \bigr) v\, D^2\, h
\,\,=\,\, -\lambda_h^{}v\,D^2\, h \,\,,
\vspace{1.5ex} \\   \displaystyle
{\cal L}_{DDH}^{} \,\,=\,\, -\bigl( \lambda_1^{}\,\cos\alpha\,\cos\beta
+ \lambda_2^{}\,\sin\alpha\,\sin\beta \bigr) v\, D^2\, H
\,\,=\,\,-\lambda_H^{}v\,D^2\,H \,\,,
\end{array}
\end{eqnarray}
with  \,$v^2=v_1^2+v_2^2$,\,  but ${\cal L}_D^{}$ has no $DDA$ term.
Since $m_0^{}$, $\lambda_1^{}$, and $\lambda_2^{}$ are all free parameters, we can treat the mass
$m_D^{}$ and the effective couplings $\lambda_{h,H}^{}$ as new free parameters in this model.

From Eq.~(\ref{yukawa_2hdm}), the Yukawa interactions of $h$ and $H$ are described by
\begin{eqnarray}
{\cal L}_{ff\cal H}^{} &=& -\bar{\cal U}_{L} M^u {\cal U}_R^{}\,
\Biggl(\frac{\cos\alpha}{\sin\beta}\,\frac{h}{v}
       + \frac{\sin\alpha}{\sin\beta}\,\frac{H}{v}\Biggr)
- \bar{\cal D}_L^{} M^d {\cal D}_R^{}\, \Biggl(-\frac{\sin\alpha}{\cos\beta}\,\frac{h}{v}
+ \frac{\cos\alpha}{\cos\beta}\,\frac{H}{v}\Biggr)
\nonumber\\ &&
-\,\, \bar E_L^{} M^l E_R^{}\, \Biggl(-\frac{\sin\alpha}{\cos\beta}\,\frac{h}{v}
+ \frac{\cos\alpha}{\cos\beta}\,\frac{H}{v}\Biggr)  \,\,+\,\, {\rm H.c.}  \,\,.
\label{yukawa}
\end{eqnarray}
We have not written down the couplings of the $CP$-odd Higgs boson $A$ to fermions because it
does not couple to $D$ and consequently $A$ is irrelevant to our darkon relic-density study,
not contributing to darkon annihilation.

To evaluate the darkon annihilation rate, we also need the couplings of $h$ and $H$ to
the weak bosons \,$V=W^\pm,Z$.\,  From the kinetic sector of the THDM, we have~\cite{thdm}
\begin{eqnarray}
{\cal L}_{VV\cal H}^{} \,\,=\,\,
\Biggl(\frac{2m_W^2}{v}\,W^{+\mu}W_\mu^-+\frac{m_Z^2}{v}\,Z^\mu Z_\mu^{}\Biggr)
\bigl[h\,\sin(\beta-\alpha)+H\, \cos(\beta-\alpha)\bigr]  \,\,.
\end{eqnarray}

\subsection{Darkon-nucleon elastic cross-section in THDM\,II+D}

Using the formulas given above, one can express the cross-section of the darkon-nucleon
elastic scattering in the THDM\,II+D as
\begin{eqnarray} \label{cs_el_2hdm}
\sigma_{\rm el}^{} \,\,\simeq\,\,
\frac{m_N^2\,v^2}{\pi\bigl(m_D^{}+m_N^{}\bigr)^2} \Biggl(\frac{\lambda_h^{}\,g_{NNh}^{\rm THDM}}{m_h^2}
+ \frac{\lambda_H^{}\,g_{NNH}^{\rm THDM}}{m_H^2}\Biggr)^{\!2} \,\,,
\end{eqnarray}
where, from Eq.~(\ref{gNNH}), the nucleon coupling to \,${\cal H}=h$ or $H$\, is
\begin{eqnarray} \label{gnnh_2hdm}
g_{NN\cal H}^{\rm THDM} \,\,=\,\,
\bigl(k_u^{\cal H}-k_d^{\cal H}\bigr) \frac{\sigma_{\pi N}^{}}{2 v} \,+\,
k_d^{\cal H}\, \frac{m_N^{}}{v}
\,+\, \frac{4k_u^{\cal H}-25k_d^{\cal H}}{27}\,\, \frac{m_B^{}}{v} \,\,.
\end{eqnarray}
The parameters $k_q^{\cal H}$ are read off from Eq.~(\ref{yukawa}) to be
\begin{eqnarray}  \label{ksm'}
k_u^h  \,\,=\,\, \frac{\cos\alpha}{\sin\beta} \,\,, \hspace{2em}
k_d^h \,\,=\,\, -\frac{\sin\alpha}{\cos\beta} \,\,, \hspace{1cm}
k_u^H  \,\,=\,\, \frac{\sin\alpha}{\sin\beta} \,\,, \hspace{2em}
k_d^H \,\,=\,\,  \frac{\cos\alpha}{\cos\beta} \,\,.
\end{eqnarray}

If both $h$ and $H$ contributed, the analysis would be complicated.  For concreteness, in our
numerical analysis we will neglect contributions from $H$ by requiring $\lambda_H^{}$ to be
very small or $m_H^{}$ very large.  Under this assumption, we can see that, since $k_{u,d}^h$
are free parameters, the situation in the THDM\,II+D can be very different from that in the SM+D.
Here it is possible for the terms proportional to $k_{u,d}^h$ in the Higgs-nucleon coupling
$g_{NNh}^{\rm THDM}$ to cancel. From Eq.~(\ref{gnnh_2hdm}), the cancelation condition is
\begin{eqnarray}
\frac{k_d^h}{k_u^h} \,\,=\,\, - \tan\alpha\,\tan\beta \,\,=\,\,
\frac{27\,\sigma_{\pi N}^{}+8\,m_B^{}}{27\,\sigma_{\pi N}^{}+50\,m_B^{}-54\,m_N^{}} \,\,.
\end{eqnarray}
Since $m_B^{}$ is related to $\sigma_{\pi N}^{}$ by Eq.~(\ref{m0}) and since $\sigma_{\pi N}^{}$
is not well determined, with values within the range
\,$35{\rm\,MeV}\lesssim\sigma_{\pi N}^{}\lesssim80$\,MeV\, having been quoted in
the literature~\cite{Cheng:1988cz,Gasser:1990ce,Ellis:2008hf}, the value of  $k_d^h/k_u^h$ has
a sizable uncertainty.
Nevertheless, we have checked that the main conclusion of this section below does not change for
this $\sigma_{\pi N}^{}$ range.
For definiteness, in the following we employ  \,$\sigma_{\pi N}^{}=45$\,MeV\,~\cite{Cheng:1988cz},
as in Eq.~(\ref{g_nnh}), yielding
\begin{eqnarray}
\frac{k_d^h}{k_u^h} \,\,\simeq\,\, -0.405 \,\,.
\end{eqnarray}

We have found that one can get the darkon-nucleon cross-section to be near the experimental
bounds by allowing  \,$\tan\alpha\tan\beta$\,  to deviate from the cancelation point, without
violating the relic density constraint. For illustration, we choose \,$\tan\alpha\tan\beta=0.45$\,
and consider two sets of $\tan\alpha$ and $\tan\beta$ values satisfying this choice:
\,$(\tan\alpha,\tan\beta)=(0.45,1)$\, and \,$(\tan\alpha,\tan\beta)=(0.45/30,30)$.\,
We note that, since low values of $\tan\beta$ are disfavored for low values of the charged-Higgs
mass $m_{H^\pm}$~\cite{WahabElKaffas:2007xd}, we will assume a large $m_{H^\pm}$
({\footnotesize$\gtrsim$}\,250\,GeV).
We also note that \,$\tan\beta=1$ and 30\, roughly span the range allowed by various
experimental and theoretical constraints~\cite{WahabElKaffas:2007xd}.

\begin{figure}[ht]
\medskip
\includegraphics[width=6.5in]{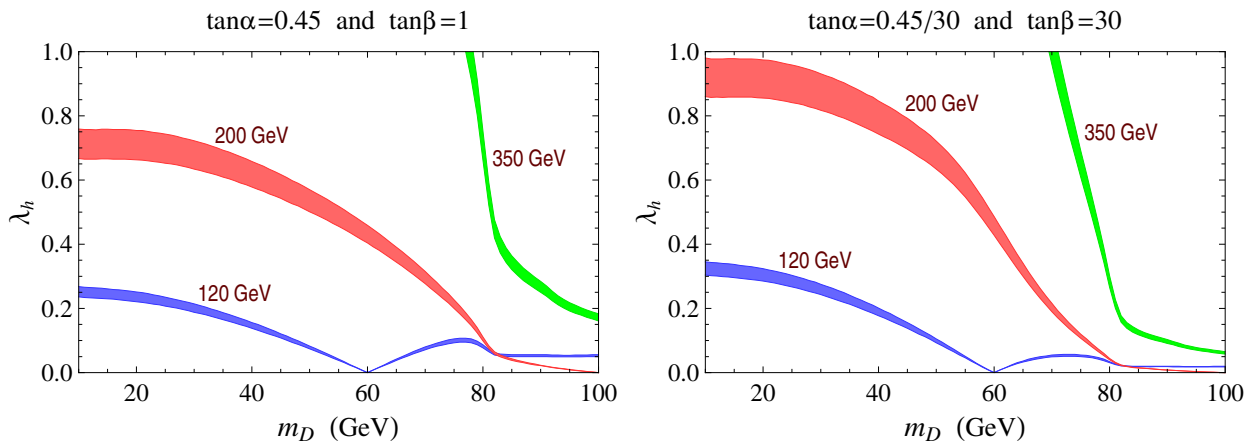}
\caption{Darkon-Higgs coupling $\lambda_h^{}$ in the THDM\,II+D as  a function of darkon mass
$m_D^{}$ for Higgs mass values  \,$m_h^{}=120,200,350$\,GeV\,  and  two cases with different
$\tan\alpha$ and $\tan\beta$ values satisfying  \,$\tan\alpha\,\tan\beta=0.45$.\,
Only regions corresponding to \,$\lambda_h^{}<1$\, have been plotted.
\label{lambda_2hdm}}
\end{figure}

\begin{figure}[ht]
\includegraphics[width=3.45in]{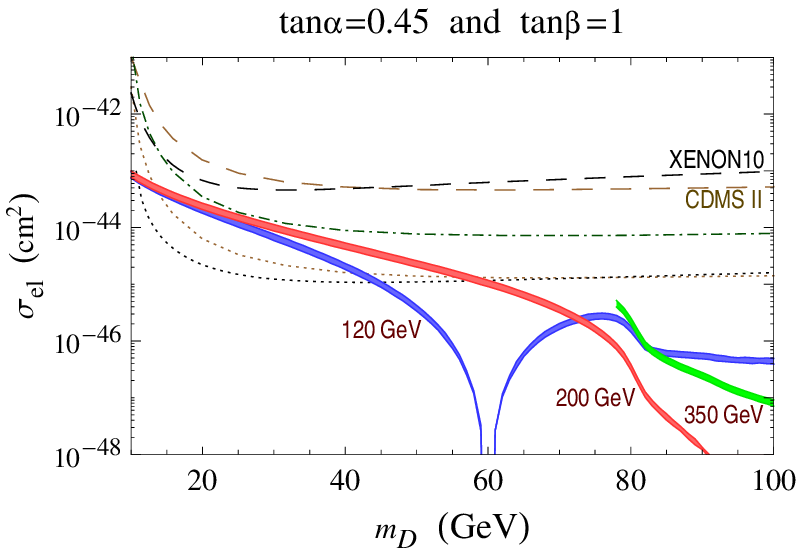}
\includegraphics[width=3.45in]{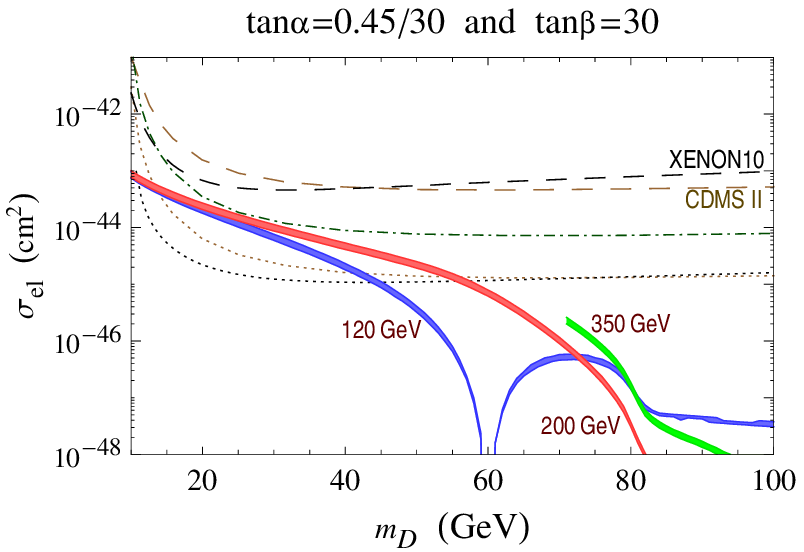}
\caption{Darkon-nucleon elastic cross-section $\sigma_{\rm el}^{}$ in the THDM\,II+D as a function
of darkon mass $m_D^{}$ for Higgs mass values  \,$m_h^{}=120,200,350$\,GeV\,  and the two cases
in Fig.~\ref{lambda_2hdm}, compared to the 90\%\,C.L. upper limits from XENON10 (black dashed curve)
and CDMS\,II (gray [brown] dashed curve), as well as projected sensitivities of SuperCDMS at Soudan (green
dot-dashed curve), SuperCDMS at Snolab (gray [brown] dotted curve), and XENON100 (black dotted curve).
\label{elastic_2hdm+d}}
\end{figure}

Since the relic density in this case is determined by the interaction of the darkon with $h$
alone, the coupling $\lambda_h^{}$ can be extracted following the steps taken in Sec.~\ref{sm+d}.
We show the resulting values of $\lambda_h^{}$ in Fig.~\ref{lambda_2hdm}.
With $\lambda_h$ determined, we can calculate the darkon-nucleon elastic cross-section,
which is now given by\footnote{It
is worth remarking that, although the $\alpha$ and $\beta$ values chosen to make
$g_{NNh}^{\rm THDM}$ sufficiently small may at the same time cause $g_{NNH}^{\rm THDM}$ to be
enhanced, the contribution of $H$ to $\sigma_{\rm el}^{}$ in Eq.~(\ref{cs_el_2hdm}) can still be
neglected, as $\lambda_H^{}$ remains a~free parameter of the model.
As Eq.~(\ref{lambda}) shows, $\lambda_{h,H}^{}$ depend on the free parameters $\lambda_{1,2}^{}$,
and so fixing $\lambda_h^{}$ does not imply fixing $\lambda_H^{}$.
Consequently, the product  \,$\lambda_H^{}\,g_{NNH}^{\rm THDM}$\, can always be made as small
as desired. In addition, as mentioned above, $m_H^{}$ can be taken to be very large to suppress
the second term in Eq.~(\ref{cs_el_2hdm}) further.}
\begin{eqnarray}
\sigma_{\rm el}^{} \,\,=\,\, \frac{m_N^2\,v^2}{\pi\,\bigl(m_D^{}+m_N^{}\bigr)^2}
\Biggl(\frac{\lambda_h^{}\,g_{NNh}^{\rm THDM}}{m_h^2}\Biggr)^{\!2} \,\,.
\end{eqnarray}
In Fig.~\ref{elastic_2hdm+d} we plot $\sigma_{\rm el}^{}$ for our parameter choice above and
\,$\lambda_h<1$.\,
We note that, as in the SM+D case, the dip of each of the \,$m_h^{}=120$\,GeV curves in this
figure at \,$m_D^{}=60$\,GeV\, corresponds to the minimum of the denominator in
Eq.~(\ref{csan}) at \,$m_h^{}=2m_D^{}$.\,
In Fig.~\ref{elastic_2hdm+d} we also plot the XENON10 and CDMS\,II upper limits, along with
the expected sensitivities of a number of future experiments~\cite{dmplot}.
Evidently, there is parameter space in the THDM\,II+D that can escape the current
experimental constraints.
Future direct-search experiments can place more stringent constraints on the model.

\section{Discussions and Conclusions}

\begin{figure}[t]
\includegraphics[width=6.7in]{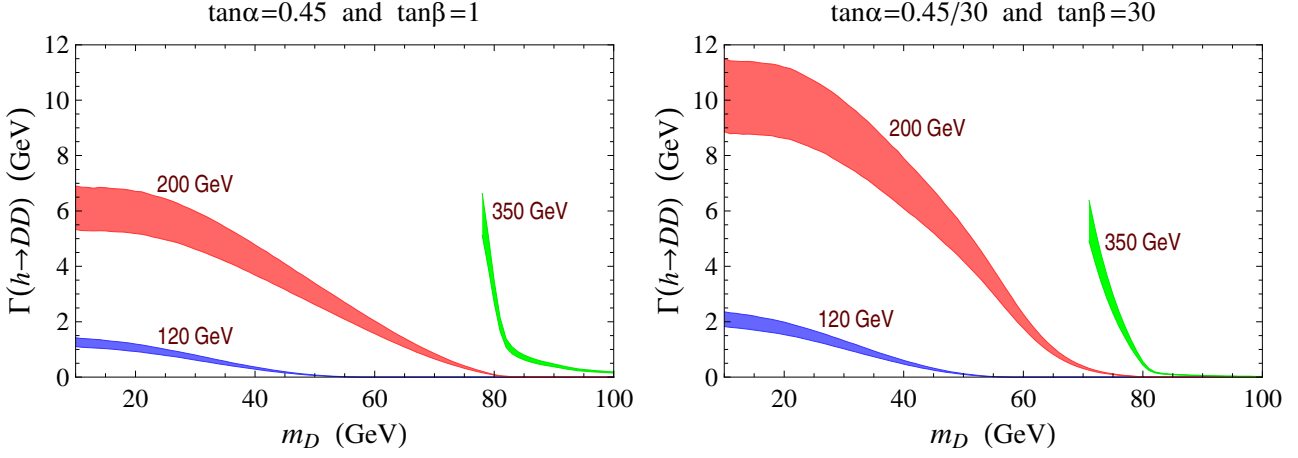}
\caption{Partial width of \,$h\to DD$\, in the THDM\,II+D as a function of darkon mass $m_D^{}$
for Higgs mass values  \,$m_h^{}=120,200,350$\,GeV\,  and the two cases in Fig.~\ref{lambda_2hdm}.
Only regions corresponding to \,$\lambda_h^{}<1$\, have been plotted.
\label{wh2dd}}
\end{figure}

\begin{figure}[t]
\includegraphics[width=6.7in]{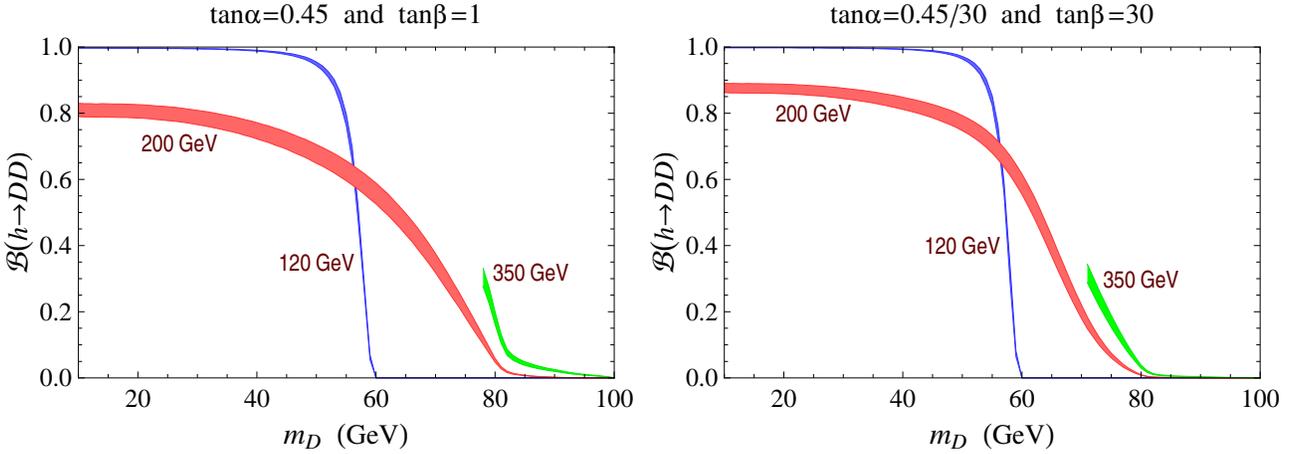}
\caption{Branching ratio of \,$h\to DD$\, in the THDM\,II+D as a function of darkon mass
$m_D^{}$,  corresponding to  \,$\Gamma(h\to DD)$\, in Fig.~\ref{wh2dd}.
\label{bh2dd}}
\end{figure}

The darkon model can have significant effects on collider physics.  Since the darkon
interacts primarily with Higgs bosons, its greatest impact is on the Higgs sector.
In either the SM+D or the THDM\,II+D, the existence of the darkon can give rise to huge enhancement
of the Higgs width via the additional process \,$h\to DD$\, because the darkon-Higgs coupling
can be sufficiently large, as can be seen from Fig.~\ref{lambda_sm} or~\ref{lambda_2hdm}.
In particular, our parameter choice above for the THDM\,II+D leads to the \,$h\to DD$\, partial
width and branching ratio shown in Figs.~\ref{wh2dd} and~\ref{bh2dd}.
We see that for small $m_D^{}$ values, the Higgs width is increased considerably and dominated
by the \,$h\to DD$\, mode, especially if \,$m_h^{}<2m_W^{}$,\, in which case the Higgs partial
width into SM particles is small.  Since the darkon is stable, the darkon pairs produced in
\,$h\to DD$\, will be invisible, and so in this case the invisible branching fraction of $h$
can be substantially bigger than its branching fraction to SM particles.
For large $m_D^{}$ values, the \,$h\to DD$\, contribution becomes small and the $h$ decay is
like that in the case without the darkon.

Although the increase of the invisible widths of Higgs bosons will obviously affect Higgs
searches at the LHC~\cite{Burgess:2000yq,He:2007tt,Barger:2007im}, it will still be able to
provide an environment for probing the darkon model.
For a Higgs boson with a large invisible branching fraction ($>$\,60\%) and a mass within the range
\,$120{\rm\,GeV}\lesssim m_h^{}\lesssim300$\,GeV,\, direct searches at the Compact Muon Solenoid 
(CMS) experiment through the usual SM modes may be unfeasible with 30\,fb$^{-1}$ of integrated 
luminosity~\cite{Barger:2007im}.
However, such a Higgs boson can be observed at ATLAS with the same integrated luminosity,
via weak-boson fusion or $Z$-Higgstrahlung, by looking for missing energy from the
decay~\cite{Barger:2007im,Davoudiasl:2004aj}.
Considering one of our THDM\,II+D examples, with \,$\tan\alpha=0.015$\, and \,$\tan\beta=30$,\,
we find that $h$ has SM-like couplings to the weak bosons and invisible branching ratios higher
than 0.6 in much of the $m_D^{}$ range, as can be seen from Fig.~\ref{bh2dd}, which serves to
illustrate the testability of the darkon model at the LHC.
Moreover, in the SM+D, if \,$m_h^{}>2m_Z^{}$,\, say \,$m_h^{}=300$\,GeV,\, the total Higgs width
can be measured with a precision of up to 10\% at ATLAS with 300\,fb$^{-1}$ integrated
luminosity~\cite{atlas}, and then the darkon contribution can be inferred after a~comparison with
the SM prediction for the width.

In conclusion, we have studied the experimental constraints from XENON10 and CDMS\,II on
the darkon model of dark matter.  In the SM+D, a darkon with mass in the ranges of 10\,GeV to
\,$(50, 70, 75)$\,GeV\,  for Higgs boson masses of  \,$(120, 200, 350)$\,GeV\,  is ruled out as
a WIMP candidate by the experimental restrictions.
By extending the SM+D to a two-Higgs-doublet model plus a~darkon, the THDM\,II+D, the experimental
limits can be circumvented due to suppression of the darkon-nucleon elastic cross-section at some
values of \,$\tan\alpha\,\tan\beta$.\,
Future DM search experiments can further constrain the parameter space of the model.
Using the darkon-Higgs coupling extracted from the DM relic density, we have found that the total
decay width of $h$ in either the SM+D or the THDM\,II+D can be greatly enhanced by a large
contribution from the invisible mode \,$h\to DD$.\,
The substantial increase of the invisible decay width of the Higgs boson would lead to
a sizable reduction of its branching fraction to SM particles.
Although this could significantly affect Higgs searches at the LHC, we expect that it will still
be able to probe the darkon model.
\\

\acknowledgments \vspace*{-1ex} 
This work was partially supported by NSC, NCTS, and NNSF.

\bigskip 

{\it Note added.}~~After this paper was accepted for publication, we became aware of a recent 
direct-search experiment by the TEXONO Collaboration~\cite{Lin:2007ka} which reported
an upper limit on the spin-independent WIMP-nucleon elastic cross section for WIMP-mass values 
below 10\,GeV, to as low as 3\,GeV. 
We have subsequently applied our analysis to darkon-mass values from 3 to 10~GeV and
found that the two models we considered, the SM+D and the THDM\,II+D, 
satisfy the bounds from TEXONO.
However, their upcoming ultralow energy germanium detector (ULEGe) experiment may be sensitive 
enough to place constraints on the models.

\appendix

\section{Derivation of Higgs-nucleon coupling\label{gnnh}}

In the $t$-channel Higgs-mediated darkon-nucleon scattering, the energy exchanged is small
compared to the nucleon mass. Therefore, we can use chiral perturbation theory at leading
order to derive the Higgs-nucleon coupling.

The relevant Lagrangian is written down in terms of the lightest meson- and baryon-octet
fields, collected into $3\times3$ matrices $B$ and $\xi$.  At leading order in the $m_s^{}$
expansion, the chiral strong Lagrangian can be written as~\cite{xpt}
\begin{eqnarray}   \label{Ls}
{\cal L}_{\rm s}^{} &=&
- m_B^{} \left\langle \bar B B \right\rangle
+ b_D^{} \left\langle\bar B\left\{M_+,B\right\}\right\rangle
+ b_F^{} \left\langle\bar B\left[M_+,B\right]\right\rangle
+ b_0^{} \left\langle M_+ \right\rangle \left\langle \bar B B \right\rangle
\nonumber \\ && \! +\,\, \mbox{$\frac{1}{2}$} f^2 B_0^{} \left\langle M_+ \right\rangle
\,\,+\,\,  \cdots \,\,,
\end{eqnarray}
where only the relevant terms are shown, $m_B^{}$ is the baryon mass in the chiral limit,
$\,\langle\cdots\rangle\equiv{\rm Tr}(\cdots)\,$ in flavor-SU(3) space, $f$ is the pion
decay constant, \,$M_+^{}=\xi^\dagger M_q^{}\xi^\dagger+\xi M_q^\dagger\xi$\, with
$\,M_q={\rm diag}\bigl(m_u^{},m_d^{},m_s^{}\bigr),\,$ and $b_{D,F,0}^{}$ and $B_0^{}$ are
free parameters which can be fixed from data.
In the following, we will take the isospin-symmetric limit $\,m_u^{}=m_d^{}=\hat m$.\,

From (\ref{Ls}), one can derive the nucleon mass $m_N^{}$ and the pion-nucleon sigma term
$\sigma_{\pi N}^{}$,
\begin{eqnarray} \label{mn}
\begin{array}{c}   \displaystyle
m_N^{}  \,\,=\,\,
m_B^{}-2\bigl(b_D^{}+b_F^{}+2b_0^{}\bigr)\hat m-2\bigl(b_D^{}-b_F^{}+b_0^{}\bigr)m_s^{} \,\,,
\vspace{2ex} \\   \displaystyle
\sigma_{\pi N}^{} \,\,=\,\, \hat m\,\partial m_N^{}/\partial\hat m \,\,=\,\,
-2\bigl(b_D^{}+b_F^{}+2b_0^{}\bigr) \hat m \,\,,
\end{array}
\end{eqnarray}
as well as expressions for the masses of the other hadrons $(\Sigma,\Xi,K,\pi)$ in terms of
$b_{D,F,0}^{}$ and $B_0^{}$.
Empirically, $\sigma_{\pi N}^{}$ is not very precisely determined and hence will be a source
of uncertainty in our calculation.
For definiteness, we adopt \,$\sigma_{\pi N}^{}=45$\,MeV\,~\cite{Gasser:1990ce}.
From the expressions for the masses and $\sigma_{\pi N}^{}$, one can obtain
\begin{eqnarray} \label{m0}
m_B^{} \,\,=\,\,
-\sigma_{\pi N}^{}\, \frac{2m_K^2+m_\pi^2}{2m_\pi^2} \,+\,
\frac{\bigl(m_\Xi^{}+m_\Sigma^{}\bigr)\bigl(2m_K^2-m_\pi^2\bigr)-2m_N^{}\,m_\pi^2}
     {4\bigl(m_K^2-m_\pi^2\bigr)} \,\,,
\end{eqnarray}
where numerically $m_K^{}$, $m_\pi^{}$, $m_\Xi^{}$, $m_\Sigma^{}$, and $m_N^{}$ are
isospin-averaged values.

We can now derive the light-quark contribution to $g_{NN\cal H}^{}$ in Eq.~(\ref{gnnh0}).
Since the terms containing $M_q^{}$ in Eq.~(\ref{Ls}) are the chiral realization of the
light-quark mass terms in the quark Lagrangian
\,${\cal L}_q^{}=-m_u^{}\,\bar u u-m_d^{}\,\bar d d-m_s^{}\,\bar s s+\cdots$,\, we have
\begin{eqnarray} \label{light}
\langle N|k_u^{}\,m_u^{}\,\bar uu+k_d^{}\,m_d^{}\,\bar dd+k_d^{}\,m_s^{}\,\bar ss|N\rangle &=&
-\langle N|k_u^{}\, m_u^{}\, \frac{\partial{\cal L}_{\rm s}^{}}{\partial m_u^{}}
+ k_d^{}\,m_d^{}\, \frac{\partial{\cal L}_{\rm s}^{}}{\partial m_d^{}}
+ k_d^{}\, m_s^{}\, \frac{\partial{\cal L}_{\rm s}^{}}{\partial m_s^{}}|N\rangle
\nonumber \\ &\!\!=&\!\!
\Bigl[ \mbox{$\frac{1}{2}$}\bigl(k_u^{}-k_d^{}\bigr)\,\sigma_{\pi N}^{} \,+\,
k_d^{}\,\bigl(m_N^{}-m_B^{}\bigr) \Bigr]\,\bar NN
\end{eqnarray}
using Eq.~(\ref{mn}), where the last line has been obtained by averaging over the $p$ and $n$
matrix elements.

For the heavy-quark contribution to $g_{NN\cal H}^{}$, we use the relation~\cite{Shifman:1978zn}
\begin{eqnarray} \label{hq}
m_Q^{}\, \bar Q Q \,\,=\,\,
-\frac{2}{3}\,\,\frac{\alpha_{\rm s}^{}}{8\pi}\, G_{a\mu\nu}^{}G_a^{\mu\nu}
\end{eqnarray}
at leading order in the heavy quark expansion, for  \,$Q=c$, $b$, or $t$,\, and the nucleon
matrix element of the trace of the energy-momentum tensor~\cite{Donoghue:1992dd}
\begin{eqnarray} \label{mn'}
m_N^{}\, \bar N N \,\,=\,\, \langle N| \theta_\mu^\mu |N\rangle  \,\,=\,\,
\langle N| m_u^{}\,\bar uu+m_d^{}\,\bar dd+m_s^{}\,\bar ss \,-\,
\frac{9\alpha_{\rm s}^{}}{8\pi}\, G_{a\mu\nu}^{}G_a^{\mu\nu}|N\rangle \,\,,
\end{eqnarray}
where  $G_a^{\mu\nu}$ is the field strength tensor of the gluons.
Setting \,$k_u^{}=k_d^{}=1$\, in Eq.~(\ref{light}) and comparing it with Eq.~(\ref{mn'}),
we conclude that
\begin{eqnarray}
\langle N| m_Q^{}\, \bar Q Q |N\rangle \,\,=\,\,
-\frac{2}{27}\, \langle N|\frac{9\alpha_{\rm s}^{}}{8\pi}\,G_{a\mu\nu}^{}G_a^{\mu\nu}|N\rangle
\,\,=\,\, \frac{2}{27}\, m_B^{}\, \bar N N \,\,.
\end{eqnarray}
It follows that
\begin{eqnarray} \label{heavy}
\langle N|k_u^{}\,m_c^{}\, \bar cc+k_d^{}\,m_b^{}\,\bar bb+k_u^{}\,m_t^{}\,\bar tt|N\rangle
\,\,=\,\, \frac{2}{27} \bigl( 2 k_u^{}+k_d^{}\bigr)\, m_B^{}\, \bar N N \,\,.
\end{eqnarray}

Combining Eqs.~(\ref{light}) and (\ref{heavy}), we finally find
\begin{eqnarray}
g_{NN\cal H}^{} \,\,=\,\,
\bigl(k_u^{}-k_d^{}\bigr) \frac{\sigma_{\pi N}^{}}{2 v} \,+\,
k_d^{}\, \frac{m_N^{}}{v} \,+\, \frac{4k_u^{}-25k_d^{}}{27}\,\, \frac{m_B^{}}{v}  \,\,.
\end{eqnarray}
We remark that, for $k_u^{}$ and $k_d^{}$ being comparable in size, $g_{pp\cal H}^{}$ and
$g_{nn\cal H}^{}$ differ from each other and from the isospin-averaged $g_{NN\cal H}^{}$ above
by at most a few percent.

\end{document}